\documentclass[prd,nofootinbib,
twocolumn,
superscriptaddress,showpacs,amsmath,amssymb]{revtex4-2}
\usepackage{amsfonts}
\usepackage{bm}
\usepackage{verbatim}
\usepackage{color}
\usepackage{graphicx}

\usepackage{subcaption}
\usepackage[title,titletoc]{appendix}

\begin{document}

\title{Emergent gravity in superplastic crystals}

\author{M.A.~Zubkov \footnote{On leave of absence from NRC "Kurchatov Institute" - ITEP, B. Cheremushkinskaya 25, Moscow, 117259, Russia}}
\affiliation{Physics Department, Ariel University, Ariel 40700, Israel}

\date{\today}

\begin{abstract}
We discuss emergent gravity in the superplastic crystals.   We restrict ourselves by the consideration of  fermions coupled to gravity. Gravity itself is either Riemannian (without torsion) or teleparallel (no curvature). We demonstrate that in this case the off - diagonal components of stress - energy tensor,  being integrated over the whole volume, are topological invariants. In equilibrium their values are not changed if the system is modified smoothly. 
  \end{abstract}
\pacs{
}
\newcommand{\rev}[1]{{#1}}

\newcommand{\rv}[1]{{#1}}

\newcommand{\rrv}[1]{\textcolor{red}{#1}}

\maketitle

\section{Introduction}

In the scenario of emergent gravity the quantum theory of gravity appears as the low energy approximation to a certain microscopic theory. Such a theory is not invariant under the general coordinate transformations. This invariance, therefore, is to appear dynamically at low energies. One of the first papers on emergent gravity has been written by A.D. Sakharov \cite{Sakharov1967}. Even earlier it was known that the defects of the crystal lattices (disclinations and dislocations) give rise to emergent Riemann - Cartan geometry, which serves as the background for the motion of various excitations  \cite{Bilby1956,Kroener1960,DzyalVol1980,KleinertZaanen2004,Kleinert2009}. However, in the real materials the dynamical theory of disclinations and dislocations does not have the form of the gravitational theory. Invariance under the general coordinate transformations  would appear \cite{Kleinert1987} if the elastic deformations do not require energy. Such systems may be called the superplastic crystals, and they were used recently to model the quantum gravity \cite{KlinkhamerVolovik2019} (see also \cite{NissinenVolovik2018a,NissinenVolovik2018b}).

There exist many other scenarios of emergent gravity. Relation of the AdS/CFT correspondence to the concept of emergent gravity has been discussed, for example, in  \cite{Ramsdonk2010}. Emergence of Einstein equations from thermodynamics has been discussed in \cite{Jacobson,Padmanabhan}. In \cite{Verlinde} gravity has been linked to entropy. The possible emergence of Einstein equations from the quantum entanglement has been considered in \cite{Lee,Swingle,Oh}.

Typically in the models with emergent gravity the huge value of cosmological constant appears. The appearance of such a value is the content of the so - called cosmological constant problem. Namely, in the quantum field theory the
vacuum fluctuations give the ultraviolet divergent contributions to the energy of vacuum. Taking into account the finite value of the ultraviolet cutoff (of the order of the Plank mass) we arrive at the value of the cosmological constant that is many orders of magnitude larger than its observed value \cite{Martin}. Calculation of the vacuum energy in quantum field theory may be found in \cite{Zeldovich}. Several scenarios that may, possibly, solve the cosmological constant problem have been proposed in the past (see, for example, \cite{Solution1,Solution2,Solution3,Solution4,Solution5,Solution6,Solution7,Solution8} and references therein). However, so far the complete solution has not been found.

In the present paper we consider the emergent gravity scenario in the model of the superplastic crystal \cite{Kleinert1987,KlinkhamerVolovik2019}. This scenario allows us to look at the Standard Model fermions as at the fermionic excitations in a superplastic crystal. The emergent torsion and curvature are caused by the dislocations and the disclinations. Their dynamics gives rise to the gravitational action. At low energies it may receive the form of the Einstein action of general relativity.  On the technical side, we will use the Wigner - Weyl formalism \cite{1,2,berezin,6} adapted in \cite{Z2016_1,FZ2019,SZ2018,ZK2017,KZ2018,chernodub2017scale,zhang2020influence,zhang2019hall} to the quantum field theory.

We discuss in the present paper, in particular, the topological invariants composed of the stress - energy tensor. The topological invariants in general reveal intimate relation between solid state systems and high energy physics  \cite{zubkov2018momentum,zubkov2012momentum,volovik2017standard,zubkov2012momentum,zubkov2017topology}. They are relevant for description of various nondissipative transport effects including chiral separation effect \cite{zubkov2023effect}, chiral vortical effect \cite{abramchuk2018anatomy} and quantum Hall effect \cite{selch2025non}. In quark matter, where the interactions are essentially non - perturbative, various topological defects dominates dynamics  \cite{bakker1999central,bakker2005standard}. Relation between relativitic quantum field theory and  fermionic superfluids has been discussed, in particular, in  \cite{volovik2013nambu,volovik2015scalar}. The two dimensional materials also reveal the intimate links with the relativistic quantum field theory \cite{katsnelson2013euler}.

The paper is organized as follows. In Section \ref{elastic} we recall the description of emergent gravity in the superplastic crystals. In Section \ref{wigner} we describe the Wigner - Weyl calculus, which allows us to prove that the traceless components of the fermionic stress energy tensor are topological invariants. In Sect. \ref{concl} we end with the conclusions.

\section{Elastic deformations and emergent gravity}

\label{elastic}

\subsection{Elastic deformations and stress tensor}

Let us parameterize the unperturbed lattice of the given crystal by coordinates $x_k, k = 1,.., D$. We will consider here the three ($D=3$) dimensional system. The displacement of each point has three components $u_a(x)$, where $a=1,2,3$. The resulting coordinates of the crystal lattice sites are
\begin{eqnarray}
y_k(x)& = & x_k + u_k(x), \quad k = 1,2,3
\end{eqnarray}
In the absence of the displacements, when $u_a=0$, the crystal is unperturbed. Metric of elasticity theory is given by
\begin{eqnarray}
h_{ik} &=& \delta_{ik} + 2 u_{ik},~~u_{ik} = \frac{1}{2}\Bigl(\partial_i u_k +
\partial_k u_i +  \partial_i u_a \partial_k u_a\Bigr),\\
&& \quad a = 1,2,3, \quad i,k = 1,...,D
\label{Two_deformations}
\end{eqnarray}

The stress tensor $T^{ij}_{elastic}$ is defined as the response of the thermodynamical potential to elastic deformation.
$$
T^{ij}_{elastic}(x) = - \frac{\delta}{ \delta u_{ij}(x)} \, {\rm log}\, Z
$$

\subsection{Elasticity tetrads}

The alternative way to describe elastic deformations is given in terms of the elasticity tetrads $E^{~a}_\mu(x)$, which represent the hydrodynamical variables of the elasticity theory \cite{DzyalVol1980,NissinenVolovik2018a,NissinenVolovik2018b}.

The three crystallographic planes may be chosen in the three dimensional ($D=3$) crystals\footnote{Those crystallographic planes are not necessarily orthogonal for the unperturbed crystals.}. The coordinates $X^a$  ($a = 1,..., D$) are introduced in such a way, that the planes given by equations $X^a(x)=2\pi n^a$, $n^a \in \mathbb{Z}$ with $a=1,...,D$ represent the mentioned crystallographic planes. In the presence of elastic deformations the chosen crystallographic planes become curved. The corresponding displacement vectors satisfy the following equation
$$
X^a(x+u)=2\pi n^a
$$
The intersections of the $D$ surfaces
\begin{equation}
X^1({\bf r},t)=2\pi n^1 \,\,, \,\,  ... \,, \,\, X^D({\bf r},t)=2\pi n^D \,,
\label{points}
\end{equation}
coincide with the positions of certain crystal lattice sites (not necessarily all of them may be found in this way). Those intersection points constitute the rectangular lattice
\begin{align}
L = \{ \mathbf{r} = {\bf R}(n_1,...,n_D) \vert \mathbf{r}\in \mathbb{R}^D, n^a \in \mathbb{Z}^D\}.
\end{align}
In order to describe dynamics we define all quantities as dependent on  imaginary time $t$. (We assume from the very beginning that the Wick rotation to space - time of Euclidean signature is performed.) The extra dimensionless coordinate $X^{D+1} = t/\Delta t$ is defined as the imaginary time in the units of $\Delta t$.
The elasticity tetrads  are gradients of the phase function
\begin{equation}
E^{~a}_\mu(y)= \partial_\mu X^a(y),\, \rev{a = 1,...,D+1}
\label{reciprocal}
\end{equation}
The inverse matrix to $E^a_\mu$ is denoted by $e_a^\mu$. For $a=1,.., D$ those quantities have units of crystal momentum, $[E^{~a}_\mu]=1/[l]$, $[e^{\mu}_a]=[l]$ while for $a = D+1$ the quantity $E^a$ has the dimension $1/[t]$.

In the absence of dislocations, the tetrads $E^a = E^{\ a}_{\mu} dy^{\mu}$ are exact differentials. They can be expressed in general form in terms of a system of deformed crystallographic coordinate planes\rev{, i.e. surfaces} of constant phase $X^a(y)=2\pi n^a$, $n^a \in \mathbb{Z}$ with $a=1,...,D$.

In the absence of dislocations, tensor $E^{~a}_\mu(y)$ satisfies the integrability condition:
\begin{equation}
 d E^a = \frac{1}{2}(\partial_\mu E^{~a}_\nu(y)-\partial_\nu E^{~a}_\mu(y)) d x^{\mu} \wedge d y^{\nu}=0\,.
\label{integrability}
\end{equation}
In the effective theory of fermions that experience vierbein $E_\mu^a$ the effective distance between the two points with coordinates $x$ and $x+dx$ is given by $|E^a_\mu dx^\mu|$. 
Matrix $e_a^\mu$ inverse to $E^a_\mu$ represents the inverse vielbein. It determines the distance in real space along the curve $\cal C$ that connects the two points in space - time of effective theory corresponding to the two values of $X$ ($X_1,X_2$) as follows:
$$
\rho_{\cal C}(X_1,X_2) = \int_{\cal C} \sqrt{e_a^\mu e_b^\nu \delta_{\mu\nu}dX^a dX^b}
$$
where $\delta_{\mu\nu} = {\rm diag}\,(1,1,1,1)$.
In dimensionless coordinates $X$ metric of real space is given by
$$
h_{ab}(X) = e_a^\mu e_b^\nu \delta_{\mu\nu} = \frac{\partial y^\mu}{\partial X^a} \frac{\partial y^\nu}{\partial X^b} \delta_{\mu\nu}
$$
It is useful to introduce the time displacement  $u^{D+1} \equiv 0$.

This allows to express metric as
$$
h_{ab}(X) =  \Big([e^{(0)}]^\mu_a + \frac{\partial u^\mu}{\partial X^a}\Big)\Big([e^{(0)}]^\nu_b + \frac{\partial u^\nu}{\partial X^b}\Big) \delta_{\mu\nu}
$$
where
$$
[e^{(0)}]^\mu_a = \frac{\partial x^\mu}{\partial X^a} = \Big([E^{(0)}]^{-1}\Big)^\mu_a
$$
and
$$
[E^{(0)}]_\mu^a = \frac{\partial X^a}{\partial x^\mu}
$$
Metric in coordinates $X$ is related to metric in coordinates $x$ as follows:
$$
\frac{\partial X^a}{\partial x^\mu}\frac{\partial X^b}{\partial x^\nu}h_{ab}(X) = h_{\mu\nu}(x) = \delta_{\mu\nu} + 2u_{\mu\nu}
$$
that is
$$
\Big[E^{(0)}\Big]^a_\mu \Big[E^{(0)}\Big]^b_\nu h_{ab}(X) = h_{\mu\nu}(x) = \delta_{\mu\nu} + 2u_{\mu\nu}
$$
Recall also, that in coordinates $y$ metric is given simply by
$$
h^{(0)}_{\mu\nu}(y) = \delta_{\mu\nu}
$$
Metric in effective theory differs from metric $h^{(0)}_{\mu\nu}(y)$. It is given by
\begin{equation}
	g_{\mu\nu} = E_\mu^a E_\nu^b \delta_{ab} = \partial_\mu X^a \partial_\nu X^b \delta_{ab}
\end{equation}
If the unperturbed crysallographic coordinates are denoted by $X^{(0)a}$, then $\partial_\mu X^a = \partial_\mu X^{(0)a}-\partial_\nu X^{(0)a} \partial_\mu u^\nu$. In the absence of dislocations and disclinations we have
\begin{eqnarray}
	g_{\mu\nu} &=&(\partial_\mu X^{(0)a}-\partial_\rho X^{(0)a} \partial_\mu u^\rho) (\partial_\nu X^{(0)b}-\partial_\sigma X^{(0)a} \partial_\nu u^\sigma ) \delta_{ab} \nonumber\\ &=& \delta_{\mu\nu} - 2 \tilde{u}_{\mu\nu},
\end{eqnarray}
where
\begin{equation}
	\tilde{u}_{ik} =  \frac{1}{2}\Bigl(\partial_i u_k +
	\partial_k u_i -  \partial_i u_a \partial_k u_a\Bigr)
\end{equation}
The response of the thermodynamical potential to the elasticity tetrads gives rise to the tensor $\Theta^{i}_a$:
$$
\Theta^{i}_a(x) = - \frac{\delta}{ \delta E_{i}^a(x)} \, {\rm log}\, Z
$$
In the similar way
$$
\Theta_{i}^a(x) = - \frac{\delta}{ \delta e^{i}_a(x)} \, {\rm log}\, Z
$$
Those tensors are related to the ordinary stress - energy tensor $T$ as follows:
$$
T^{ij} = \frac{1}{2} e^{\{j}_b \Theta^{i\}}_a \delta^{ab}, \quad T_{ij}=\frac{1}{2} g_{k\{j}\Theta_{i\}}^a e^{k}_a
$$
(Notice that $T_{ij}$ differs from elastic stress tensor.)


\subsection{Teleparallel gravity}

\rev{In the presence of the single dislocation (placed at the line $\xi$) the coordinates $X$ become ambiguous. The vielbein $E_\mu^a$ being integrated along the closed path $\cal C$ surrounding $\xi$ gives the Burgers vector specific for the given lattice and the given type of the dislocation \footnote{Its value is quantized according to the lattice type}:
\begin{equation}
\int_{\cal C} dy^\mu E_\mu^a = B^a \label{burgers}
\end{equation}
In the differential form this equation is given by
$$
\partial_\mu E_\nu^a(y) - \partial_\nu E_\mu^a(y) = B^a \epsilon_{\mu\nu\rho}\int d\xi^{\rho}(\tau) \delta^{(3)}(y-\xi(\tau)) , \, D = 3
$$
The dislocations distributed continuously give the vielbein $E^a_\mu$ that (being averaged) corresponds to torsion
$$
T_{\mu\nu}^{..a}(y) = \partial_\mu E_\nu^a(y) - \partial_\nu E_\mu^a(y)
$$
We may speak of the field $T_{\mu\nu}^a$ as of the continuous function of coordinates. }

\rev{The curved geometry given by the vielbein $E_\mu^a$ is, actually, the geometry of the teleparallel gravity. This is the version of Riemann - Cartan gravity with vanishing curvature (and spin connection), and nonzero torsion. The corresponding space is parallelizable Weitzenbock space. The possible terms in the action of the dynamical theory of this type of gravity may be found in \cite{Diakonov:2011fs}. If we restrict ourselves by the terms with the derivatives up to the second, and if there is no selected direction in $D+1$ dimensional space - time, while the P and T symmetries hold, then we are left with
\begin{equation}
S = \int d^{D+1}y \, {\rm det} \, E \Big(\lambda  + \alpha v^2 + \beta a^2 + \gamma t^2 \Big)\label{ST}
\end{equation}
Here $a,v,t$ are the irreducible components of torsion:
$$
a^\kappa = \frac{1}{4 \, {\rm det}\,E}\epsilon^{\kappa \alpha \beta \gamma}T_{\alpha \beta \gamma}
$$
$$
v_\kappa = T_{\kappa \nu}^{..\nu}
$$
$$
t_{\mu\nu}^{..\lambda} = T_{\mu\nu}^{..\lambda} + \delta^\lambda_{[\mu}T_{\nu]\rho}^{..\rho} - g^{\lambda \rho}T_{\rho [\mu \nu]}
$$
For certain particular values of $\alpha, \beta, \gamma$ we will obtain the so called teleparallel equivalent of general relativity. }


We consider the quantum vacuum as the spacetime superplastic crystal, where $X^a(y)=2\pi n^a$, $n^a \in \mathbb{Z}$ with $a=1,2,3,4$ is the system of the four  deformed crystallographic coordinate hyperplanes in $3+1$ D.
One may assume that in  the superplastic vacuum, the effective action for $E^{~a}_\mu$ is the gravitational \rev{one given by Eq. (\ref{ST}). Metric $g_{\mu\nu}$, which originates from the elasticity tetrads, is:
\begin{equation}
g_{\mu\nu}=\delta_{ab}E^{~a}_\mu E^{~b}_\nu\,.
\label{metric}
\end{equation}}


\subsection{Riemann - Cartan geometry}

\rv{Above we discussed the case, when the torsion is present while the curvature is absent. This occurs in the presence of dislocations. The disclinations give rise to the  curvature. This occurs as follows. The disclination corresponds to the removal of a sector ended at the given line. The remaining parts of the lattice are glued together. Therefore, the path around the given line captures the angle different from $2\pi$. This means that the curvature is nonzero. In the unperturbed lattice each path corresponds to vanishing parallel transporter. In the presence of a disclination the parallel transporter along the closed path, which encloses the disclination, is equal to the matrix of rotation. The distribution of disclinations may be described by continuous  $SO(3)\subset SO(4)$ connection.  This connection modifies derivative as usual: it becomes covariant. In the absence of dislocations we deal, therefore, with  Riemannian geometry.}

\rv{If both the disclinations and the dislocations are present, there is the emergent Riemann - Cartan geometry. The basic variables are the vielbein $E^a_\mu$, and the spin connection $\omega_{\mu}\in so(D+1)$, where $D=3$. The mentioned above parallel transporter around the disclination results in the rotation matrix:
\begin{equation}
P\,{\rm exp}\,\Big(\int_{\cal C} \omega_\mu dy^\mu\Big) = e^{\phi {\rm l}\hat{\Sigma}}\label{Sigma_P}
\end{equation}
Here $\phi$ is the angle specific for the given disclination, while $\rm l$ is the unity vector along the axis of rotation, while $\hat \Sigma$ is the generator of rotation. As usual, curvature $R_{\mu\nu}$ is defined as
$$
{\rm lim}_{|A|\to 0}P\,{\rm exp}\,\Big(\int_{\partial A} \omega_\mu dy^\mu\Big) = {\rm exp}\,\Big(\frac{1}{2}\int_{A} R_{\mu \nu} dy^\mu \wedge dy^\nu \Big)
$$
The covariant derivative is given by
$$
D_\mu = \partial_\mu + \omega_\mu
$$
while the definition of torsion is modified as follows:
$$
T_{\mu\nu}^{..a}(y) = D_\mu E_\nu^a(y) - D_\nu E_\mu^a(y)
$$
There are a lot of possible local terms in the action that contain derivatives up to the second. Those terms may be composed in different way of torsion, curvature, and various covariant derivatives of the vielbein. These terms were classified in \cite{Diakonov:2011fs}.}

\subsection{The fields experienced by the fermions in the presence of dislocations and disclinations}

{Let us consider the tight - binding model with Hamiltonian
\begin{eqnarray}
H = \sum_{X_1 X_2} \bar{\Psi}^A_{X_1} t^{AB}_{X_1 X_2} \Psi^B_{X_2}
\end{eqnarray}
Hopping parameters $t^{AB}_{X_1 X_2}$ correspond to the jumps of the fermions between the sites $X_1$ and $X_2$ of the lattice. The simplest dependence of the hopping parameters  on the elastic deformations is
\begin{eqnarray}
&& t^{AB}_{X_1 X_2} = t^{AB} + t^{AB}\beta \Big(\sqrt{e^\mu_a e^\nu_b \delta_{\mu\nu} (X_1^a - X_2^a)(X_1^b - X_2^b)}\nonumber\\&&-\sqrt{[e^{(0)}]^\mu_a [e^{(0)}]^\nu_b \delta_{\mu\nu} (X_1^a - X_2^a)(X_1^b - X_2^b)}\Big)
\end{eqnarray}
Here $\beta$ is the so - called Gruneisen parameter. Suppose that the model admits the topologically protected Fermi points. In their vicinities the emergent Weyl fermions appear (this is the case of the so - called Weyl semimetals). In the absence of multiple dislocations and disclinations the above mentioned modification of the hopping parameters results in the appearance of  emergent gauge field and  emergent teleparallel gravity. The emergent gauge field appears in the first approximation to the complete theory, while the emergent gravity appears in the next approximation. The emergent gravity is characterized by torsion while the emergent gauge field is characterized by its field strength. The presence of multiple dislocations and disclinations modifies the emergent gravity experienced by the fermions. Namely, it has been shown that the dislocations carry the emergent singular  torsion flux and the emergent singular $U(1)$ flux. In the similar way one may demonstrate, that the disclinations carry the singularity of curvature. }

{The especially interesting case is when the Gruneisen parameter $\beta$ is small $\beta \ll 1$ \footnote{Notice that in graphene, for example, $\beta \sim 2$.}. In this situation we deal with the Weyl fermions that experience {\it the described  above gravity with torsion}. For the superplastic crystal those elastic deformations, which are not caused by the appearance of the disclinations and the dislocations, do not require energy. {\it As a result, the low energy effective theory appears to be invariant under the diffeomorphisms.}  } {In the low energy effective theory the distance between the points is understood as the distance in the lattice units, while the coordinates of lattice sites are understood as parametrization. Modifying the distance between lattice sites we perform reparametrization. Distance between infinitely close points with coordinates $x$ and $x+dx$ in the effective theory is given by $|E_\mu^a dx^\mu|$.}

Thus, we may suppose that the Standard Model with gravity appears as the low energy description of the superplastic crystal with the fermionic excitations inside it. The real gravity appears as the theory of the elastic vielbein. The presence of the dislocations results in torsion while the presence of disclinations results in curvature. Notice, that the mentioned here tight - binding model may be thought of as an effective model, while the true microscopic theory remains continuous. {\bf The "real" coordinates $x^\mu$ now represent the parametrization of space, while the  distance in effective theory is measured in cristallographic coordinates $X^a$.}

In this pattern of space - time the time still remains distinguished from space. The curvature and the torsion do not have nontrivial time components. The description of the space - time gravity and the Standard Model fermions as an effective theory may be thought of as the theory written in the reference frame with $E^{D+1}_\mu = \delta^{D+1}_{\mu}\frac{1}{\Delta t}$. This is the theory with the partially fixed  gauge. (The symmetry with respect to the space - time diffeomorphisms is reduced down to the  diffeomorphisms of space.)

There is an alternative way to look at the problem of the distinguished (imaginary) time direction. Notice, that $E^a_\mu$ enters all expressions in such a way, that there is no essential difference between the components with $a,\mu = 1,2,3$ and $a$ (or $\mu$) equal to $4$. We may extend the notion of the elastic deformations (and the corresponding displacement vectors $u_\mu$) from the $3$ - component vectors defined in $3$ - dimensional space to the $4$ - component vectors defined in $4$ - dimensional space - time. We may actually think that space - time is the four - dimensional superplastic crystal. This restores the complete symmetry between space and (imaginary) time components of all vectors and tensors. The dislocations will appear as the two - dimensional defects, such that for the closed contours that enclose them, the integrals of Eq. (\ref{burgers}) are nontrivial. The disclinations also appear as the two - dimensional defects. The integrals over the contours that surround them given by Eq. (\ref{Sigma_P}) are nontrivial. $\Sigma$ represents the generators of the $SO(4)$ group (recall that we perform the Wick rotation to the Euclidean signature of space - time).

\section{The theory on the language of Wigner - Weyl formalism}

\label{wigner}

{\it From now on we consider the effective low energy theory of the superplastic crystal. In this theory the coordinates are continuous, and they represent the parametrization. These coordinates are inherited from the "real" coordinates of points in the superplastic crystal.} In principle we can forget about the superplastic crystal, and simply consider the theory  described by the partition function
\begin{equation}
	Z = \int D \bar{\Psi} D \Psi e^{ \int d^{D+1} y (\bar{\Psi} \hat{Q}\Psi ) }
\end{equation}
Operator $\hat{Q}$ contains dependence on coordinates and momentum, and the dependence on coordinates completely determined by dependence of $\hat{Q}$ on the vierbein $E_\mu^a(x)$. {\bf  We impose on the theory the requirement that it is invariant under the reparametrizations.} 

\subsection{Wigner - Weyl formalism}

In this section we consider the Wigner - Weyl formalism to be used for the description of the theory. For more details of this formalism we refer to \cite{Z2016_1,ZK2017,SZ2018,FZ2019}.
\rv{Let us start from the discussion of the model with the noninteracting fermions in $D+1$ space - time dimensions. It is useful to introduce  the (Grassmann - valued) distribution function:
$$
W(x,y) =\langle x|\hat{W}|y\rangle = {\bar{\Psi}(x)}{\Psi}(y)
$$
we also denote the matrix elements of operator $\hat Q$ by
$$
Q(x,y) = \langle x|\hat{Q}|y\rangle
$$
As a result, the partition function may be rewritten as follows
\begin{equation}
Z = \int D \bar{\Psi} D \Psi e^{  {\bf Tr}\, \hat{W} \,\hat{Q} }
\end{equation}
Here in the exponent the functional trace appears.
The next step is the introduction of the Weyl symbol of operator $\hat A$:
\begin{equation} \begin{aligned}
{A}_W(p,x) \equiv \int dy e^{-iy p} {A}({x+y/2}, {x-y/2})\label{GWx}
\end{aligned} \end{equation}
We introduce the Weyl symbols of operators $\hat Q$ and $\hat W$, and come to
\begin{eqnarray}
Z &=&
\int D\bar{\Psi}D\Psi \, {\rm exp}\Big( \int d^{D+1}x \int \frac{d^{D+1}p}{(2\pi)^{D+1}} \nonumber\\ &&{\rm Tr} {W}_W[\Psi,\bar{\Psi}](p,x) \star Q_W(p,x) \Big)\label{dZ2}
\end{eqnarray}
We denote here by the star (Moyal) product the following expression
$$
* = e^{\frac{i}{2} \left( \overleftarrow{\partial}_{x}\overrightarrow{\partial_p}-\overleftarrow{\partial_p}\overrightarrow{\partial}_{x}\right )}
$$}

\subsection{Derivative of the partition function as a topological invariant}

Let us suppose that operator $\hat Q$ introduced above depends on a parameter $\eta$. We consider derivative of the logarithm of the partition function:
\begin{eqnarray}
&& \frac{d}{d\eta} {\rm log}\,Z = \frac{1}{Z} \int D\bar{\Psi} D\Psi \Big(  \int d^{D+1}x \int \frac{d^{D+1}p}{(2\pi)^{D+1}} \nonumber\\ &&{\rm Tr} {W}_W[\psi,\bar{\psi}](p,x)\star Q_W(p,x) \Big) \Big(  \int d^{D+1}x \int \frac{d^{D+1}p}{(2\pi)^{D+1}} \nonumber\\ &&{\rm Tr} {W}_W[\psi,\bar{\psi}](p,x) \star \frac{d}{d\eta} Q_W(p,x) \Big)\nonumber\\ &&=
 \int d^{D+1}x \int \frac{d^{D+1}p}{(2\pi)^{D+1}} {\rm Tr} {G}_W(p,x) \star \frac{d}{d\eta} Q_W(p,x) \label{dZe}
\end{eqnarray}
As it was mentioned above, we made here rotation to space - time of Euclidean signature, i.e. we consider the theory in imaginary time.
Here $Q_W$ depends on parameter $\eta$, and we introduced the Wigner transformed Green function $G_W$ as the Weyl symbol of $\hat G$, where
$$
G(x,y) = \langle x|\hat{G}|y\rangle =\frac{1}{Z} \int D \bar{\Psi} D \Psi e^{  {\bf Tr}\, \hat{W} \,\hat{Q} }\bar{\Psi}(x)\Psi(y)
$$
As usual in field theory, in order to calculate this integral in practise we need to put the system into the large, but finite spatial volume. Correspondingly, the imaginary time extent (the inverse temperature) also should be large, but finite. Then, strictly speaking, the sums over the discrete momenta appear instead of the integral. Though, we are able to calculate the integrals instead of the sums because the overall volume is large. Correspondingly, the boundary conditions are to be imposed. The experience with the conventional quantum field theory prompts that for the large volume (and the large imaginary time extent) the type of the boundary condition is irrelevant for the physics in the bulk. For the definiteness we assume the periodical spacial boundary conditions, and the anti - periodic boundary conditions for the imaginary time direction.

The Wigner transformed Green function obeys the Groenewold equation:
$$
Q_W \star G_W = 1
$$
The properties of the star product guarantee the commutativity
\begin{eqnarray}
&&\int d^{D+1}x \int \frac{d^{D+1}p}{(2\pi)^{D+1}} {\rm Tr} {A}_W(p,x) \star B_W(p,x) \nonumber\\ &&= \int d^{D+1}x \int \frac{d^{D+1}p}{(2\pi)^{D+1}} {\rm Tr} {B}_W(p,x) \star  A_W(p,x) \nonumber\\ &&= {\bf Tr} \hat{A} \hat{B}
\end{eqnarray}
Here $\bf Tr$ is the functional trace of an operator that obeys ${\bf Tr} \hat{A} \hat{B} =  {\bf Tr} \hat{B} \hat{A}$.
Therefore, since $\delta (Q_W\star G_W) =0$, the variation of $\frac{d}{d\eta} {\rm log}\, Z$ is equal to zero:
\begin{eqnarray}
&& \delta \frac{d}{d\eta} {\rm log}\,Z =
 \int d^{D+1}x \int \frac{d^{D+1}p}{(2\pi)^{D+1}}\nonumber\\ && {\rm Tr}\Big( {G}_W(p,x) \star \frac{d}{d\eta} \delta Q_W(p,x)+\delta{G}_W(p,x) \star \frac{d}{d\eta} Q_W(p,x)\Big)\nonumber\\&&=\int d^{D+1}x \int \frac{d^{D+1}p}{(2\pi)^{D+1}} \nonumber\\ &&{\rm Tr}\Big(  {G}_W(p,x) \star \frac{d}{d\eta} \delta Q_W(p,x)\nonumber\\&&-{G}_W(p,x) \star \delta{Q}_W(p,x)\star {G}_W(p,x) \star \frac{d}{d\eta} Q_W(p,x)\Big)
 \nonumber\\&&=\int d^{D+1}x \int \frac{d^{D+1}p}{(2\pi)^{D+1}}\nonumber\\ && {\rm Tr}\Big( {G}_W(p,x) \star \frac{d}{d\eta}  \delta Q_W(p,x)\nonumber\\&&-G_W(p,x) \star \delta Q_W(p,x)\star {G}_W(p,x) \star \frac{d}{d\eta} Q_W(p,x)\Big)
 \nonumber\\&&=\int d^{D+1}x \int \frac{d^{D+1}p}{(2\pi)^{D+1}} {\rm Tr}\Big( {G}_W(p,x) \star \frac{d}{d\eta}  \delta Q_W(p,x)\nonumber\\&&+\frac{d}{d\eta} G_W(p,x) \star \delta Q_W(p,x)\Big)\nonumber\\&&=\int d^{D+1}x \int \frac{d^{D+1}p}{(2\pi)^{D+1}} \frac{d}{d\eta}  {\rm Tr} \, {G}_W(p,x) \star \delta Q_W(p,x) \label{deltaZ}
\end{eqnarray}

We come to an interesting conclusion that the derivative of  (the logarithm of) the partition function is topological invariant provided that the last integral in Eq. (\ref{deltaZ}) vanishes.

In the above proof it has been implied that the integrals are convergent. The absence of the infrared divergencies means that neither $\hat{G}$ nor $\hat{Q}$ have poles at finite values of momenta. As for the ultraviolet divergencies, we assume that the discussed  theory does not contain them. It has the form of a lattice model (the imaginary time is discretized as well). The formalism described in Sect. \ref{elastic}, which leads to the Groenewold equation, formally has been derived for the continuous theory. It remains valid for the lattice models as well if the external field (in our case this is the elastic deformation tensor) varies slowly, so that its variation at the distance of the order of the lattice spacing may be neglected (for the details see \cite{Z2016_1,SZ2018,FZ2019}). This condition is assumed in the theory of elasticity. 

Thus Eq. (\ref{dZe}) at zero temperature is topological invariant for the gapped systems that do not contain ultraviolet divergencies provided that
\begin{equation}
\int d^{D+1}x \int \frac{d^{D+1}p}{(2\pi)^{D+1}} \frac{d}{d\eta}  {\rm Tr} \, {G}_W(p,x) \star \delta Q_W(p,x) = 0\label{condition}
\end{equation}
for the given class of variations $\delta Q_W(p,x)$.
Notice, that the theories containing gapless fermions in the presence of interactions suffer from the infrared divergencies, we do not consider them here.

An example, when the derivative of partition function with respect to a parameter is  topological invariant is given by the electric current density integrated over the whole system volume. In this case the role of parameter $\eta$ is played by constant external electromagnetic potential $A_k$ along the $k$ - th axis. Then in Eq. (\ref{condition}) we substitute  $\frac{d}{d\eta}  $ by $-\frac{\partial}{\partial p_k}$, and this condition is satisfied if we choose appropriate boundary conditions in momentum space. The topological invariance of the total electric current derived in this way represents a field theoretical generalization of the quantum - mechanical Bloch theorem \cite{ZZ2019_3}.


\subsection{Contribution of fermions to the stress - energy tensor}
\label{Ttop}

Here we will consider example of the topological invariant given by the traceless components of stress - energy tensor integrated over the whole volume. In the absence of nontrivial background metric this stress tensor is given by response of partition function to the variations of elastic deformations $-\delta \tilde{u}_{ij}(x)$.
In the presence of the background metric originated from distribution of disclinations and dislocations we consider the response of thermodynamical potential to the variation of effective metric given by
$$
\delta g^{ij}(x) = - \delta u^k \partial_k g^{ij}(x) + g^{ik}(x)\partial_k  \delta u^j(x) + g^{jk}(x)\partial_k  \delta u^i(x)
$$
and effective vielbein
$$
\delta e^{i}_a(x) = - \delta u^k \partial_k e^{i}_a(x) +  e^{j}_a(x)\partial_j  \delta u^i(x)
$$
In the considered theory the response to variation of $g^{ij}$ represents stress - energy tensor $T_{ij}(x)$. Response to the variation of $e^i_a$ gives tetrad components of the stress - energy tensor $\Theta_\mu^a$. In our consideration it is more convenient to deal with the latter. Relations between those two tensors are given by
$$
\Theta_{i}^j = \Theta_i^a e_a^j, \quad T_{ij} = \frac{1}{2}(\Theta^k_i g_{kj}+\Theta^k_j g_{ki})
$$
The total stress - energy tensor is given by an integral of  $e^j_a \Theta_{i}^a(x)$ over the whole volume averaged in time $\langle e^j_a \Theta_{i}^a \rangle$:
\begin{eqnarray}
&&\langle \Theta_{i}^j \rangle = T \,\int d^{D+1}x \langle e^j_a \Theta_{i}^a(x) |E(x)|\rangle \label{Tij}\\ && =2 T\int d^{D+1}z |E(z)|
 \int d^{D+1}x \nonumber\\&&\int \frac{d^{D+1}p}{(2\pi)^{D+1}} {\rm Tr}\, {G}_{W}(p,x) \star e^j_a(z) \frac{\delta}{ \delta e^{i}_a(z)} Q_{W}(p,x)\nonumber
\end{eqnarray}
Here $T$ is temperature, which is assumed to be small. If $Q_W$ and $G_W$ do not depend on (imaginary) time, then $T\int dx^{D+1}$ becomes equal to unity. In Eq. (\ref{Tij}) we ignore the possible contribution to stress - energy tensor that may appear due to the dependence of measure over fermionic fields on $|E(x)|$. This contribution will be discussed later.

Using technique developed in Appendix A it is shown in Appendix C that we are able to substitute here $e_a^j\frac{\delta}{\delta e_{i}^a(z)}$ by $\frac{1}{2}\Big(p_j\frac{\partial}{\partial p_i}  - \delta_i^j\Big)\delta(z-x)$ provided that $e^{i}_a$ depend on coordinates only slightly. In Appendix B we show, in addition, that the covariant derivative of stress - energy tensor is a topological invariant.  


The effective low energy theory for the model with a Fermi point gives rise to the following expression for $Q_W$
\begin{equation}
{Q}_W(p,y) =  \frac{|E(y)|}{2}  e^\mu_a(y) \{\gamma^a, (p_\mu -\omega_\mu(y))\} \label{Qlow0}
\end{equation}
provided that the Gruneisen parameter is much smaller than unity. This situation may be realized in the ideal superplastic crystals, where the elastic deformations do not require energy. Here $\omega$ is the emergent spin connection resulted from the disclinations, while the elasticity vielbein leads to the presence of torsion.  As it was mentioned above, we assume that the appearance of the Fermi point is broken by a small mass term. As a result the fermions become gapped, and the system already does not suffer from the infrared singularities (which appear for any gapless fermions in the presence of interactions). Thus we come to
\begin{equation}
{Q}_W(p,y) =  \frac{|E(y)|}{2}  e^\mu_a(y) \{\gamma^a, (p_\mu-\omega_\mu(y))\}  - m |E(y)|\label{Qlow}
\end{equation}
It was noticed above that Eq. (\ref{Qlow}) represents the low energy approximation for the theory with the Fermi point (broken weakly by a small mass term). For momenta far from the Fermi point the form of $\hat Q$ is essentially different. The whole theory may be actually considered as the regularization of the relativistic theory with fermions and  gravity.

{\it Assuming the absence of dislocations we arrive at the theory of emergent Riemannian gravity, in which the spin connection is determined unambiguously by the vielbein. Alternatively we can assume that the disclinations are absent, and then we deal with the teleparallel gravity. In both cases the only source of dependence on coordinates in the theory is the vielbein depending on coordinates.}

Taking into account all mentioned above, we may try to consider the following expression as the definition of the stress - energy tensor:
\begin{eqnarray}
\langle \Theta_{i}^j \rangle &=& T \,\int d^{D+1}x \langle \Theta_{i}^j(x)|E(x)| \rangle \label{Tij2}\\ && =T
 \int d^{D+1}x |E(x)| \int \frac{d^{D+1}p}{(2\pi)^{D+1}} {\rm Tr}\, {G}_{W}(p,x) \star \nonumber\\&& \Big(p_i\frac{\partial}{\partial p_j} -  \delta_{i}^j\Big) Q_{W}(p,x)\nonumber
\end{eqnarray}
Applying this expression formally to Eq. (\ref{Qlow}) we would obtain:
\begin{eqnarray}
\langle \Theta_{i}^j \rangle &=&   T
 \int d^{D+1}x |E(x)| \int \frac{d^{D+1}p}{(2\pi)^{D+1}} {\rm Tr}\, {G}_{W}(p,x) \star \nonumber \\&& \Big(-\delta_{i}^j Q_W(p,x)  + {|E|} e^j_a \gamma^a (p_{i} - \omega_{i}(x))\Big)\label{nonregularized}
\end{eqnarray}
For the variation of stress - energy tensor we get
\begin{eqnarray}
 \langle \delta \Theta^j_{i} \rangle &=&  T
 \int d^{D+1} x |E(x)| \int \frac{d^{D+1}p}{(2\pi)^{D+1}} {\rm Tr}\,\nonumber \\&& \Big(\frac{\partial}{\partial p_j}p_i -  \delta_i^j\Big) {G}_{W}(p,x) \star \delta Q_{W}(p,x)\nonumber\\&& - \frac{i}{2} T
 \int d^{D+1} x |E(x)| \int \frac{d^{D+1}p}{(2\pi)^{D+1}} \frac{\partial}{\partial p_j}\frac{\partial}{\partial x^i}\,{\rm Tr}\,\nonumber \\&& {G}_{W}(p,x) \star \delta Q_{W}(p,x)\nonumber\\&& - T
 \int d^{D+1} x |E(x)| \int \frac{d^{D+1}p}{(2\pi)^{D+1}} {\rm Tr}\,\nonumber \\&& E_\mu^a \delta e^\mu_a  {G}_{W}(p,x) \star \Big(p_i\frac{\partial}{\partial p_j}-  \delta_i^j\Big) Q_{W}(p,x)\nonumber
\end{eqnarray}
The second term originates from the non - commutativity of $p_i$ and the Moyal product, $p_i \star f - f\, p_i = -\frac{i}{2}\partial_{x^i} f$, and is kept here up to the terms linear in the derivatives with respect to $x$. Being a total derivative in $p$, it does not contribute under the appropriate boundary conditions in momentum space.
When the gauge $|E(x)|=1$ is kept, we have $E_\mu^a \delta e^\mu_a = -\delta\,{\rm log}\,|E| = 0$, so that the last term drops out. Combining the remaining two terms with the help of the relation $p_i \star f = p_i f - \frac{i}{2}\partial_{x^i} f$, we obtain
\begin{eqnarray}
 \langle \delta \Theta^j_{i} \rangle &=&  T
 \int d^{D+1} x \int \frac{d^{D+1}p}{(2\pi)^{D+1}} {\rm Tr}\,\nonumber \\&& \Big[\frac{\partial}{\partial p_j}\Big(p_i \star {G}_{W}(p,x) \star \delta Q_{W}(p,x)\Big)\nonumber\\&& -  \delta_i^j\, {G}_{W}(p,x) \star \delta Q_{W}(p,x)\Big]\nonumber
\end{eqnarray}
One can see, that under a suitable ultraviolet regularization the traceless components of stress - energy tensor $\Theta^i_j$ (those with $i\ne j$) are topological invariants provided that $|E(x)|$ is not changed during the given variations. This may always be achieved via a suitable reparametrization that accompanies the modification of the system.

 The whole expression Eq. (\ref{nonregularized}) suffers from all possible types of divergencies - both ultraviolet and infrared. The infrared divergency is to be regularized through the finite overall volume (and finite temperature). The naive ultraviolet regularization (via the ultraviolet cutoff in the  integral over momentum) will then give the ultraviolet divergent expression, which does not have a form of topological invariant.   The superplastic crystal by itself represents an ultraviolet regularization.

\section{Conclusions}

\label{concl}

In the present paper we discuss the scenario, in which the Standard Model of elementary particles and the quantum gravity appear in the effective low energy description of a condensed matter system. The latter system describes the ideal superplastic crystal, in which smooth elastic deformations do not require energy. As a result, the emergent theory is invariant under the general coordinate transformations of space. In this theory the disclinations result in curvature, while the dislocations result in torsion. The action, which describes the dynamics of dislocations and disclinations, may have at low energies the form of the Einstein action of general relativity.

Alternatively, the superplastic crystals may be considered as the ultraviolet  regularization of quantum gravity coupled to fermions. The advantage of this regularization is that the integration measure over fermions does not depend on the gauge fields and on the gravitational fields. Instead, the dependence of operator $Q_{W}(x,p)$ on momenta is responsible for the quantum anomalies. Therefore, the variational derivative of the integration measure does not contribute the  total stress energy tensor.

We consider the two particular cases - when the gravity is either teleparallel (disclinations and spin connection are absent) or Riemannian (dislocations and torsion are absent). Then the traceless components of the fermionic stress energy tensor  are topological invariants, provided that the fermions are massive and the background metric depends on coordinates weakly.

The author is grateful for the discussions to G.E.Volovik, who actually proposed him to consider the superplastic crystals as the source for the emergent gravity. 

\appendix

\renewcommand{\theequation}{\Alph{section}.\arabic{equation}}
\setcounter{equation}{0}

\section{Variation of Weyl symbol of an operator with respect to reparametrizations}

Let us consider variation of Wigner transformed matrix elements of an operator $\hat A$ caused by the following transformation:
$$
A(y_1,y_2) \to {A}(y_1 -\delta u(y_1), y_2 -\delta u(y_2))
$$
with small elastic deformation displacement $\delta u^k(x)$:
\begin{eqnarray}
&&A_W(p,x)+\delta {A}_W(p,x)\nonumber\\&& \equiv \int dy_1 dy_2  e^{-i(y_1-y_2) p} {A}(y_1 -\delta u(y_1), y_2 -\delta u(y_2))\nonumber\\&&\delta(x-(y_1+y_2)/2)\label{GWxA}
\end{eqnarray}
This gives
\begin{eqnarray}
&&\delta {A}_W(p,x) =  \int dy_1 dy_2  e^{-i(y_1-y_2) p} \delta(x-(y_1+y_2)/2)\nonumber\\&&
\Big(-\delta u^k(y_1) \partial_{y^k_1}-\delta u^k(y_2) \partial_{y^k_2}  \Big)\nonumber\\&&{A}(y_1 , y_2 )\nonumber\\&&=
\int dy_1 dy_2  \frac{dq }{(2\pi)^4}e^{-i(y_1-y_2) p} \delta(x-(y_1+y_2)/2)\nonumber\\&&
\Big(-\delta u^k(y_1) \partial_{y^k_1}-\delta u^k(y_2) \partial_{y^k_2}  \Big)\nonumber\\&&e^{i(y_1-y_2) q}{A}_W(q , (y_1+y_2)/2)\nonumber\\
&&=
\int dy_1 dy_2  \frac{dq }{(2\pi)^4}e^{-i(y_1-y_2) p} \delta(x-(y_1+y_2)/2)\nonumber\\&&
\Big(-\delta u^k(y_1) (iq_k + \partial_{x^k}/2)-\delta u^k(y_2) (-iq_k+\partial_{x^k}/2)  \Big)\nonumber\\&&e^{i(y_1-y_2) q}{A}_W(q ,x)\nonumber\\
&&=
\int dy_1 dy_2  \frac{dq }{(2\pi)^4}e^{i(y_1-y_2)(q- p)} \delta(x-(y_1+y_2)/2)\nonumber\\&&
\Big(-\delta u^k(y_1) (iq_k + \partial_{x^k}/2)-\delta u^k(y_2) (-iq_k+\partial_{x^k}/2)  \Big)\nonumber\\&&{A}_W(q ,x)\nonumber\\
\end{eqnarray}
Therefore,
\begin{eqnarray}
\frac{\delta {A}_W(p,x)}{\delta u^k(z)} & = & \int dy_2  \frac{dq }{(2\pi)^4}e^{i(z-y_2)(q- p)} \delta(x-(z+y_2)/2)\nonumber\\&&
 (-iq_k - \partial_{x^k}/2) {A}_W(q ,x)\nonumber\\
 && + \int dy_1  \frac{dq }{(2\pi)^4}e^{i(y_1-z)(q- p)} \delta(x-(z+y_1)/2)\nonumber\\&&
 (iq_k - \partial_{x^k}/2) {A}_W(q ,x)\nonumber\\
 & = & 2^{D+1}\int  \frac{dq }{(2\pi)^4}e^{2i(z-x)(q- p)} \nonumber\\&&
 (-iq_k - \partial_{x^k}/2) {A}_W(q ,x)\nonumber\\
 && +2^{D+1} \int  \frac{dq }{(2\pi)^4}e^{2i(x-z)(q- p)} \nonumber\\&&
 (iq_k - \partial_{x^k}/2) {A}_W(q ,x)\nonumber\\
\end{eqnarray}
We assume that boundary conditions in momentum space are organized in such a way, that we may shift the integration variable $q \to q-p = k$. This may always be achieved via the appropriate ultraviolet regularization.
As a result we are able to represent the above written variational derivative as
\begin{eqnarray}
\frac{\delta {A}_W(p,x)}{\delta u^k(z)}
 & = & -2^{D+1}\int  \frac{dk }{(2\pi)^4}e^{2i(z-x)k} \nonumber\\&&
 (i(k_k+p_k)_{} + \partial_{x^k}/2) {A}_W(k+p ,x)\nonumber\\
 && - 2^{D+1}\int  \frac{dk }{(2\pi)^4}e^{2i(x-z)k} \nonumber\\&&
 (-i(k_k+p_k)_{} + \partial_{x^k}/2) {A}_W(k+p ,x)\nonumber\\
 & = & - 2^{D+1}\int  \frac{dk }{(2\pi)^4}e^{2i(z-x)k+k_j \partial_{p_j}} \nonumber\\&&
 (i(p_k)_{} + \partial_{x^k}/2) {A}_W(p ,x)\nonumber\\
 && -  2^{D+1}\int  \frac{dk }{(2\pi)^4}e^{2i(x-z)k+k_j\partial_{p_j}} \nonumber\\&&
 (-i(p_k)_{} + \partial_{x^k}/2) {A}_W(p ,x)\nonumber\\
 & = &- 2^{D+1}\int  \frac{dk }{(2\pi)^4}e^{2i(z-x)k} e^{i\overleftarrow{\partial}_{x^j} \overrightarrow{\partial}_{p_j}/2} \nonumber\\&&
 (i(p_k)_{} + \partial_{x^k}/2) {A}_W(p ,x)\nonumber\\
 && - 2^{D+1}\int  \frac{dk }{(2\pi)^4}e^{2i(x-z)k}e^{-i\overleftarrow{\partial}_{x^j} \overrightarrow{\partial}_{p_j}/2}  \nonumber\\&&
 (-i(p_k)_{} + \partial_{x^k}/2) {A}_W(p ,x)\nonumber\\
   & = & -\delta(z-x) e^{i\overleftarrow{\partial}_{x^j} \overrightarrow{\partial}_{p_j}/2} (i(p_k)_{} \nonumber\\&&+ \partial_{x^k}/2) {A}_W(p ,x)\nonumber\\
 && - \delta(z-x)e^{-i\overleftarrow{\partial}_{x^j} \overrightarrow{\partial}_{p_j}/2} (-i(p_k)_{} \nonumber\\&&+ \partial_{x^k}/2) {A}_W(p ,x)\nonumber\\
 & = & -\delta(z-x) \star (i(p_k)_{} \nonumber\\&&+ \partial_{x^k}/2) {A}_W(p ,x)\nonumber\\
 && - \delta(z-x) (-i(p_k)_{} \nonumber\\&&+ \partial_{x^k}/2) {A}_W(p ,x)\star \delta(z-x) \nonumber
\end{eqnarray}
Coming back to $\delta A_W$ we may write it in the following form:
\begin{eqnarray}
\delta A_W(p,x) &=& -i \Big(\delta u^k(x) \star  \Big[p_k  {A}_W(p ,x)\Big] \nonumber\\ && - \Big[p_k  {A}_W(p ,x)\Big] \star \delta u^k(x)\Big)\nonumber\\ &&- \frac{1}{2}  \Big(\delta u^k(x) \star \partial_{x^k} {A}_W(p ,x)\nonumber\\ && + \partial_{x^k}{A}_W(p ,x) \star \delta u^k(x)\Big)\label{AA}
\end{eqnarray}

\section{Covariant derivative of stress - energy tensor in terms of Wigner transformed Green function}

We know that with respect to reparametrizations $y\to x(y)=y-\delta u$ the action $S =  \int d^{D+1} y_1 d^{D+1} y_2 \bar{\Psi}(y_1)\langle y_1| \hat{Q}[g]| y_2 \rangle \Psi(y_2)$ is invariant:
\begin{eqnarray}
S&=& \int d^{D+1} x_1 d^{D+1} x_2 \bar{\Psi}(x_1)\langle x_1| \hat{Q}[e]| x_2 \rangle \Psi(x_2)\nonumber\\
&=& \int d^{D+1} x(y_1) d^{D+1} x(y_2) (\bar{\Psi}(x(y_1))\nonumber\\
&&\langle x(y_1)| \hat{Q}[g]| x(y_2) \rangle \Psi(x(y_2))\nonumber \\
&=&\int d^{D+1} y_1 d^{D+1} y_2 \bar{\Psi}(x(y_1))\Big|\frac{\partial x}{\partial y}(y_1)\Big|\nonumber\\
&&\langle x(y_1)| \hat{Q}[g]| x(y_2) \rangle \Big|\frac{\partial x}{\partial y}(y_2)\Big| \Psi(x(y_2))\nonumber \\
&=&\int d^{D+1} y_1 d^{D+1} y_2 \bar{\Psi}(x(y_1))\nonumber\\
&&\langle y_1| \hat{Q}[e^x]| y_2 \rangle  \Psi(x(y_2))
\end{eqnarray}
Here $e^x$ is the corresponding transformation of vielbein. Thus the reparametrizations inside $\hat{Q}$
\begin{eqnarray}
\langle y_1| \hat{Q}| y_2 \rangle &\to & \langle y_1 - \delta u(y_1)| \hat{Q}| y_2 - \delta u(y_2) \rangle \nonumber\\&& \Big|1-\frac{\partial \delta u}{\partial y}(y_1)\Big|\Big|1-\frac{\partial \delta u}{\partial y}(y_2)\Big| \label{trans}
\end{eqnarray}
compensate transformation $\Psi(y) \to \Psi(y - \delta u)$ and correspond to the transformation of vielbein. Recall that spinors are subject to $SU(2) \otimes SU(2)$ transformations that are not related to the reparametrizations of coordinates. Spin connection is transformed as vector with respect to the transformations of coordinates. However, in Riemannian geometry the spin connection is expressed through the vielbein, and therefore, its transformation is caused by the appropriate transformation of the latter. Therefore, indeed the transformation of Eq. (\ref{trans}) is reduced to the transformation of $e^i_a$. This transformation gives:
\begin{eqnarray}
Q_W \to  (1-\partial_i \delta u^i) \star (Q_W + \delta Q_W) \star (1-\partial_i \delta u^i)
\end{eqnarray}
where $\delta Q_W$ has been calculated in Appendix A.
Let us denote the resulting overall variation (that takes into account both the transformation of matrix elements of $\hat Q$ and the extra factors $\Big|1-\frac{\partial \delta u}{\partial y}\Big|$) by $\delta_2$. We get
$$
\delta_2 Q_W = \delta Q_W - \partial_i\delta u^i \star Q_W - Q_W \star \partial_i\delta u^i
$$
and
\begin{eqnarray}
\delta_2 Q_W &= & -i \Big(\delta u^k(x) \star  \Big[p_k  {Q}_W(p ,x)\Big] \nonumber\\ && - \Big[p_k  {Q}_W(p ,x)\Big] \star \delta u^k(x)\Big)\nonumber\\ &&- \frac{1}{2}  \Big(\delta u^k(x) \star \partial_{x^k} {Q}_W(p ,x)\nonumber\\ && + \partial_{x^k}{Q}_W(p ,x) \star \delta u^k(x)\Big) \nonumber\\ && - \partial_i\delta u^i \star Q_W - Q_W \star \partial_i\delta u^i \label{AAA}
\end{eqnarray}

First we demonstrate the power of the proposed technique and calculate response of the thermodynamical potential to constant $\delta u^k$. Such a $\delta u^k$ causes variation of metric equal to $\delta g^{ij} = -\delta u^k \partial_k g^{ij} $. Response of ${\rm log}\, Z$ to the variation $\delta u^k$ gives the covariant derivative of stress - energy tensor. According to Eq. (\ref{deltaZ}) the corresponding response of $Q_W(p,x)$ to $ \delta u^j = {\rm const}$ is
\begin{eqnarray}
\delta^{(0)} Q_W(p,x) &=& - \delta u^k  \partial_{x^k} {Q}_W(p ,x)\label{AA2}
\end{eqnarray}
Then
\begin{eqnarray}
&&\langle  D_i T_{ij} \rangle = T \,\int d^{D+1}x \langle D_i T_{ij}(x) \rangle \sqrt{|g(x)|}\label{TijA3}\\ && =- T
 \int d^{D+1}x \int \frac{d^{D+1}p}{(2\pi)^{D+1}} {\rm Tr}\, {G}_{W}(p,x) \star \frac{\partial}{\partial \delta u^{j}} Q_{W}(p,x)\nonumber\\ && =
 T \int d^{D+1}x \int \frac{d^{D+1}p}{(2\pi)^{D+1}} {\rm Tr}\, {G}_{W}(p,x) \star \partial_{x^j} Q_{W}(p,x)
\end{eqnarray}
According to Eq. (\ref{deltaZ}) variation of  the covariant derivative of total stress - energy tensor is given by
\begin{eqnarray}
&&\delta \langle D_i T_{ij} \rangle = T \,\int d^{D+1}x \delta \langle D_iT_{ij}(x) \rangle \sqrt{|g(x)|}\label{TijA22}\\ && =- T
 \int d^{D+1}x \int \frac{d^{D+1}p}{(2\pi)^{D+1}}\frac{\partial}{\partial \delta u^{j}}  {\rm Tr}\, {G}_{W}(p,x) \star \delta  Q_{W}(p,x)\nonumber\\ && =
T \int d^{D+1}x \int \frac{d^{D+1}p}{(2\pi)^{D+1}}  \partial_{x^j} {\rm Tr}\, {G}_{W}(p,x) \star \delta Q_{W}(p,x)
\end{eqnarray}
The last integral vanishes if we take appropriate boundary conditions in coordinate space. Therefore, the covariant derivative of stress - energy tensor integrated over the whole volume is robust to the smooth modification of the system. This means that the global Einstein anomaly  is a topological invariant.  (Recall that at $D=3$ the field theory in conventional regularization predicts no Einstein (Weyl) anomaly.)

\section{Stress - energy tensor in terms of Wigner - Weyl calculus}

Equipped with Eq. (\ref{AAA}) we derive here response of partition function to the variation of an induced vielbein.  Recall, that the latter variation has the form
$$
\delta e^{i}_a(x) =  - \delta u^k \partial_k e^{i}_a(x) + e_a^{k}(x)\partial_k  \delta u^i(x)
$$
in the presence of background vielbein $e^{i}_a$.
First we calculate the response of $\delta_2 Q_W(p,x)$ to $\partial_k \delta u^j$. This calculation does not give an immediate expression for the response to the vielbein  variation. But it allows to calculate it indirectly. In order to obtain this response we use derivative expansion in Eq. (\ref{AA}). The term proportional to the first order in derivatives gives:
\begin{eqnarray}
&& \delta^{(1)}_2 Q_W(p,x) =  \partial_{p_j} \Big(p_k {Q}_W \Big) \partial_{j}\delta u^k(x) - 2 Q_W\partial_{j}\delta u^j(x)\nonumber\\ && =   \partial_{p_j} \Big(p_k {Q}_W\Big) E^a_j e_a^l \partial_{l}\delta u^k(x)- 2 Q_W E^a_j e_a^l\partial_{l}\delta u^j(x)
\nonumber\\ && =   \Big[\partial_{p_j} \Big(p_{k} {Q}_W(p ,x)\Big)E^a_j \Big] (\delta e_a^k(x) + \delta u^m(x) \partial_m e_a^k(x)) \nonumber\\ &&  - 2 Q_W(p,x) E^a_j (\delta e^j_a(x) +  \delta u^m(x) \partial_m e_a^j(x))\label{d1}\end{eqnarray}
Let us suppose, that vielbein is given by a slight variation of the constant one $e^i_a(x) = h^i_a(x) + e^{(0)i}_{a}$, where $e^{(0)i}_{a}= {\rm const} (x)$.
We know that $\delta u^k$ may enter the expression for $\delta Q_W$ only through the variation $\delta e^i_a$. Therefore, the overall variation may be expressed through the kernel ${\cal K}$, which is to be determined:
\begin{eqnarray}
\delta_2 Q_W(p,x) &=& \int dz \, {\cal K}^{a}_{i}(p,x;z)\, \delta e^i_a(z), \nonumber\\ {\cal K}^{a}_{i}(p,x;z) &\equiv& \frac{\delta_2 Q_W(p,x)}{\delta e^{i}_a(z)}\label{kernelK}
\end{eqnarray}
The same kernel allows to express the term of the zeroth order in the derivatives of $\delta u$. {\bf At this point we assume that $Q_W(p,x)$ depends on $x$ only through the background vielbein. This means that either the spin connection is expressed explicitly trough vielbein (which is the case of Riemannian gravity) or spin connection is absent at all (teleparallel gravity, Weitzenbock geometry).} We have for $\delta^{(0)}_2 Q_W(p,x)$:
 \begin{eqnarray}
\delta^{(0)}_2 Q_W(p,x) &=& - \delta u^k  \partial_{x^k} {Q}_W(p ,x)\nonumber\\&\approx&
- \delta u^k \int dz \, \Big[\partial_{z^k} h^i_a(z)\Big]\, {\cal K}^{a}_{i}(p,x;z)\label{d0K}
\end{eqnarray}
Combining the two leading terms of the derivative expansion (given by Eq. (\ref{d1}) and Eq. (\ref{d0K})) we get up to the terms linear in $h$ and its first derivatives the following equation:
\begin{eqnarray}
&&\int dz \, {\cal K}^{a}_{i}(p,x;z)\, \delta e^i_a(z) \nonumber\\ && =  \Big[\partial_{p_j} \Big(p_{k} {Q}_W(p ,x)\Big)E^a_j \Big] (\delta e_a^k(x) + \delta u^m(x) \partial_m e_a^k(x)) \nonumber\\ &&  - 2 Q_W(p,x) E^a_j (\delta e^j_a(x) +  \delta u^m(x) \partial_m e_a^j(x))\nonumber\\&&
- \delta u^k \int dz \, \Big[\partial_{z^k} h^i_a(z)\Big]\, {\cal K}^{a}_{i}(p,x;z)\label{Keq}
\end{eqnarray}
This is the equation that determines the kernel ${\cal K}$. Notice that the unknown kernel enters both its sides.

Let us solve Eq. (\ref{Keq}). It should be satisfied for arbitrary $\delta u(x)$. Although the reparametrizations are parametrized by the $D+1$ functions $\delta u^i$, while $\delta e^i_a$ contains $(D+1)^2$ independent components, at any given point $x$ the reparametrizations generate an arbitrary variation of the vielbein. Namely, taking $\delta u(x) = 0$ with arbitrary $\partial_l \delta u^i(x)$ we obtain $\delta e^i_a(x) = e^l_a(x)\, \partial_l \delta u^i(x)$, which spans all the components of $\delta e^i_a(x)$ because the matrix $e^l_a$ is invertible. To the leading order of the derivative expansion the kernel is local:
\begin{equation}
{\cal K}^{a}_{i}(p,x;z) = M^{a}_{i}(p,x)\, \delta(x-z)\label{Kloc}
\end{equation}
We substitute Eq. (\ref{Kloc}) into Eq. (\ref{Keq}) and use the relations $\delta e^i_a = -\delta u^m \partial_m e^i_a + e^l_a \partial_l \delta u^i$, $\partial_m h^i_a = \partial_m e^i_a$, and $E^a_j e^l_a = \delta^l_j$. The left hand side of Eq. (\ref{Keq}) receives the form
$$
M^{a}_{i}\, \delta e^i_a = - M^{a}_{i}\, \delta u^m\, \partial_m e^i_a + M^{a}_{i}\, e^l_a\, \partial_l \delta u^i
$$
while the right hand side is reduced to
$$
\partial_{p_l}\Big(p_i {Q}_W\Big)\, \partial_l \delta u^i - 2\, Q_W\, \partial_l \delta u^l - M^{a}_{i}\, \delta u^m\, \partial_m e^i_a
$$
The terms proportional to $\delta u^m$ without the derivatives are identical in the both sides, and cancel each other for any $M^{a}_{i}$. (This was to be expected: the undifferentiated $\delta u$ corresponds to the translation of the background, while the translation of $Q_W$ is given by the translation of the vielbein substituted to the kernel.) The remaining terms are proportional to $\partial_l \delta u^i$, which is an arbitrary matrix at the given point $x$. Therefore,
$$
M^{a}_{i}\, e^l_a = \partial_{p_l}\Big(p_i {Q}_W\Big) - 2\, \delta^l_i\, {Q}_W
$$
Multiplying the both sides by the inverse matrix $E$, and using $\partial_{p_l}\Big(p_i {Q}_W\Big) = \delta^l_i {Q}_W + p_i \partial_{p_l} {Q}_W$, we obtain $M^{a}_{i} = \Big[E^a_k p_{i}\partial_{p_k} - E^a_i\Big]{Q}_W$, that is
\begin{eqnarray}
&& \frac{\delta_2 Q_W(p,x)}{\delta e^{i}_a(z)}
 =  \Big[E_k^a p_{i}\partial_{p_k} - E^a_i\Big] {Q}_W(p ,x) \delta(x-z)\end{eqnarray}
The total stress - energy tensor receives the form
\begin{eqnarray}
\langle e_a^j(x) \Theta^a_{i}(x) \rangle &=&  T
 \int d^{D+1}x d^{D+1}z |E(z)|\nonumber\\&&\int \frac{d^{D+1}p}{(2\pi)^{D+1}} {\rm Tr}\, {G}_{W}(p,x) \star e_a^j \frac{\delta}{ \delta e^{i}_a(z)} Q_{W}(p,x)\nonumber\\ & = &
T \int d^{D+1}x |E|\int \frac{d^{D+1}p}{(2\pi)^{D+1}} {\rm Tr}\, {G}_{W}(p,x)\nonumber\\&& \star \Big[p_i\partial_{p_j} - \delta_i^j\Big] Q_{W}(p,x)\nonumber
\end{eqnarray}
This expression being understood naively is divergent both in ultraviolet and in infrared. Infrared regularization goes through the consideration of the system in large but finite volume. The superplastic crystal by itself represents an ultraviolet regularization. It is worth mentioning, that metric is a quantity, which is well defined in continuum theory. Correspondingly the stress energy tensor being the response of the system to the variation of metric, is also well - defined in continuum theory only.

\section{Functional logarithm}

In the main text the functional Logarithm is used. It enters expression for the partition function $Z \equiv {\rm Det} \hat{Q} = {\rm exp} \, {\bf Tr} \,{\rm Log}\, \hat{Q}$.
One can also define this logarithm using a suitably regularized representation:
\begin{eqnarray}
{\bf Tr}\,O\,({\rm Log}\,\hat{Q} - {\rm Log}\,\hat{Q}_0)&=& -{\rm lim}_{\epsilon \to 0} \int_\epsilon^\infty \frac{ds}{s} \int d^{D+1}x \nonumber\\&&\langle x| O e^{i s \hat{Q}} - O e^{i s \hat{Q}_0} |x \rangle \label{OL}
\end{eqnarray}
Here $O$ is a certain operator. Convergence of the last expression is guaranteed if all eigenvalues of $\hat Q$ are real, and we add to each of them the small imaginary part $i\delta$, which gives the mentioned above regularization. Thus it is assumed, that after the Wick rotation $\hat Q$ is Hermitian. Under such a regularization the integral over $s$ becomes convergent at $s \to \infty$.

One can check that Eq. (\ref{OL}) is manifestly not invariant under the reparametrizations. One can see this expanding the exponent as a series in powers of $\hat Q$. An example of $\hat Q$ given by Eq. (\ref{Qlow}) shows how the invariance under the reparametrizations is broken due to the product of determinants $|E(x)|$. In order to have the expression invariant under the reparametrizations we, therefore, have to fix the gauge $|E(x)| = 1$. Then, Eq. (\ref{OL}) gives the proper definition of the functional logarithm in this gauge. It has the same value for the configurations related to the given one by reparametrizations, where already $|E(x)| \ne 1$.

Thus dealing with Eq. (\ref{OL}) we assume the above mentioned gauge fixing. If $O$ commutes with $\hat Q$ and $\delta \hat{Q}$, one can easily calculate (in the gauge $|E(x)|=1$)
\begin{eqnarray}
\delta {\bf Tr}\,\hat{O}\,{\rm Log}\,\hat{Q} & =& -i\int d^{D+1}x \langle x|\int_0^\infty {ds}  \,O\,e^{i s \hat{Q}}\delta \hat{Q}| x \rangle \nonumber\\&=& -i\int_0^\infty {d\bar{s}} \int d^{D+1}x e^{i \bar{s}}\langle x|\hat{O} \hat{Q}^{-1}\delta \hat{Q}|x\rangle
\nonumber\\&=&  \int d^{D+1}x \langle x|\hat{O} \hat{Q}^{-1}\delta \hat{Q}|x\rangle\nonumber\\
& =& \int d^{D+1}x \frac{d^{D+1}p}{(2\pi)^{D+1}}{O}_W \star G_W \star \delta {Q}_W
\end{eqnarray}
For $O_W = O(p)$ that is the function of $p$ only the last expression may be easily generalized to its covariant form adding the extra factor $|E(x)|$:
\begin{eqnarray}
\delta {\bf Tr}\,\hat{O}\,{\rm Log}\,\hat{Q} & =&  \int d^{D+1}x|E(x)| \frac{d^{D+1}p}{(2\pi)^{D+1}}{O}(p) \nonumber\\&&G_W \star \delta {Q}_W
\end{eqnarray}

\end{document}